\PassOptionsToPackage{bookmarksnumbered,unicode}{hyperref}
\RequirePackage{hyperref}
\documentclass[manuscript,screen,nonacm]{acmart}
\usepackage{wrapfig}
\usepackage{subfigure}

\AtBeginDocument{%
  \providecommand\BibTeX{{%
    \normalfont B\kern-0.5em{\scshape i\kern-0.25em b}\kern-0.8em\TeX}}}

\copyrightyear{2023}
\acmYear{2023}
\setcopyright{rightsretained}
\acmConference[AutomotiveUI '23 Adjunct]{15th International Conference on Automotive User Interfaces and Interactive Vehicular Applications}{September 18--22, 2023}{Ingolstadt, Germany}
\acmBooktitle{15th International Conference on Automotive User Interfaces and Interactive Vehicular Applications (AutomotiveUI '23 Adjunct), September 18--22, 2023, Ingolstadt, Germany}\acmDOI{10.1145/3581961.3609868}
\acmISBN{979-8-4007-0112-2/23/09}

\begin{document}

\title{Exploring Mixed-Reality for Enhancing Driver Warning Systems: A Preliminary Study on Attention-Shifting Methods and Hazard Perception}

\renewcommand{\shorttitle}{Exploring Mixed-Reality for Enhancing Driver Warning Systems}


\author{Shih-Yu Ma}
\authornote{Both authors contributed equally to this research.}
\email{m10910301@mail.ntust.edu.tw}
\orcid{0000-0001-6686-0079}
\affiliation{%
  \institution{National Taiwan University of Science and Technology}
  \city{Taipei}
  \country{Taiwan (R.O.C)}
}

\author{Nolan Robert Brady}
\authornotemark[1]
\email{Nolan.Brady@colorado.edu}
\orcid{0009-0005-0863-3659}
\affiliation{%
  \institution{University of Colorado Boulder}
  \city{Boulder}
  \state{Colorado}
  \country{USA}
}

\author{Xu Han}
\email{xuha2442@colorado.edu}
\orcid{0000-0003-3928-8227}
\affiliation{%
  \institution{University of Colorado Boulder}
  \city{Boulder}
  \state{Colorado}
  \country{USA}
}

\author{Neng-Hao Yu}
\email{jonesfish@gmail.com}
\orcid{0000-0003-0717-8269}
\affiliation{%
  \institution{National Taiwan University of Science and Technology}
  \city{Taipei}
  \country{Taiwan (R.O.C)}
}

\author{Tom Yeh}
\email{tom.yeh@colorado.edu}
\orcid{0000-0003-0227-9781}
\affiliation{%
  \institution{University of Colorado Boulder}
  \city{Boulder}
  \state{Colorado}
  \country{USA}
}

\renewcommand{\shortauthors}{Ma and Brady, et al.}

\begin{abstract}
Modern vehicles are equipped with advanced sensors and capabilities, yet engaging human drivers effectively in hazard perception remains a challenging research area. This paper investigates the potential of mixed reality (MR) to enhance driver warning systems. The study examines the impact of attention-shifting methods, including visual cues, audio cues, and a combination of both, on drivers' hazard perception. A preliminary study involving six participants was conducted, and the NASA-TLX analysis did not yield significant differences. However, through the use of functional near-infrared spectroscopy (fNIRS), we unveil distinct brain activation patterns associated with visual and sound cues. Further research with larger sample sizes and diverse driving scenarios is required to validate and expand upon these preliminary results.

\end{abstract}





\maketitle


\par\medskip
\noindent
\begin{minipage}{\linewidth}
\footnotesize

\noindent
This is the author's version of the work.

\medskip

\noindent
The definitive version was published in the
Adjunct Proceedings of the 15th International
Conference on Automotive User Interfaces and
Interactive Vehicular Applications
(AutomotiveUI '23 Adjunct),
pp. 87--92, 2023.

\medskip

\noindent
DOI:
\href{https://doi.org/10.1145/3581961.3609868}%
{10.1145/3581961.3609868}

\end{minipage}
\par\smallskip


\section{Introduction}
Driving requires constant processing of external information, which must be presented to the driver in a clear and easily understandable way to ensure safety, especially during critical situations. Current systems like Head-Up Displays (HUDs) and AR windshields (WSD) have limitations in displaying information within a confined area and focal depth, potentially missing important alerts outside this restricted region.

In contrast, Mixed Reality (MR) devices have the advantage of projecting digital information anywhere in the driver's field of view, providing more customizable and adaptable visualizations. This ensures that information is presented in the most relevant and convenient location for the driver. MR devices also offer different focal depth displays and introduce interactive elements, which can dynamically direct the driver's attention to sudden events.

However, the impact of this emerging technology on drivers' cognitive load is not yet fully understood. To address this research gap, our study aims to investigate how drivers shift their attention during critical situations when using mixed-reality displays. We will employ functional near-infrared spectroscopy (fNIRS) and self-report methods like NASA-TLX to comprehensively analyze data and shed light on the potential benefits and challenges of MR displays in driving scenarios.
\section{Related work}
\paragraph{Driver's Situation Awareness}
Situation Awareness (SA) plays a crucial role in driving, encompassing perception, comprehension, and projection of environmental elements~\cite{283529}. SA in driving refers to operators' updated and meaningful knowledge guiding their decisions and actions~\cite{RN487}. Given the attention-demanding nature of driving, maintaining adequate SA is essential. In related research, gaze behavior, including fixation time and dwell time, has been utilized to detect user awareness of environmental hazards~\cite{RN488, kim2020toward, martin2018dynamics, RN494}. These studies have shown the potential of gaze behavior as an indicator of hazard detection. In our study, we employed gaze information to used as a measurement tool to establish reaction time.

\paragraph{Attention Guidance in Driving Scenario}
With recent advances in automotive technology, different in-vehicle display systems have emerged to provide drivers with surrounding information~\cite{RN486, chen2014design, song2018lane}. A significant research focus in this domain centers around enhancing drivers' SA through the effective design of notifications \cite{RN486}. Traditional displays like HUDs and WSDs are able to project notifications onto a small windshield area~\cite{guo2014analyzing, 8466859, gabbard2014behind, haeuslschmid2016design, riegler2019augmented}. 

However, their fixed focal depth and projected notification position raise challenges. Prior research, like Coleman et al.'s study on comparing different HUD displays \cite{8466859}, emphasized the importance of addressable focal planes in minimizing reaction time and improving reaction precision~\cite{8466859}. Due to the limited field of view(FOV) and testing scenarios they didn't further discuss attention shifting in their research.
In our paper we addresses the attention shifting gap by exploring the potential benefits and challenges of mixed reality displays for attention guidance in driving scenarios.

\paragraph{Brain sensing with fNIRS}
fNIRS has emerged as a valuable tool for non-invasive monitoring of brain activities. Similar to fMRI, it estimates brain activity by measuring brain hemodynamic responses. By using light sources and detectors on the scalp, fNIRS captures the absorption rate of light to determine the concentration and oxygenation of blood in the cerebral cortex. This allows for objective reflection of activation patterns in specific brain areas\cite{saikia_validation_2021}. fNIRS is portable, reliable, and widely used in studying cognitive and task switching\cite{chance1998novel, hyafil_two_2009, japee_role_2015}. It offers flexibility, lower costs, and minimal disruption from small movements and physiological signals. Validation against fMRI confirms its reliability, although fNIRS has limited spatial resolution and anatomic specificity compared to fMRI due to its light-based sensing technique.

\section{Attention Shifting Test}
\subsection{Study overview} 
Our study focuses on investigating the workload associated with attention shifting methods with MR displays in driving scenarios, utilizing fNIRS as a measuring tool. We developed a driving simulator with MR displays and implemented three attention-shifting methods. Through a within-subjects experiment (n=6), we compared brain activity and behavioral performance across these methods. By analyzing fNIRS signals, self-reported surveys, and behavioral metrics, we aim to address the following research questions:\\
RQ1: How does the type of notification impact cognitive workload, specifically in terms of cortical responses measured by fNIRS?\\
RQ2: How does the type of notification affect response time and error rate in a driving context?\\
We present detail of our methodology and result in the following sections.

\paragraph{Participants}
Six participants (20-27 years old, average age 24.5) were recruited from the university. They had normal vision and the experiment was conducted in a campus research laboratory.

\paragraph{Task}
The study task is divided into a primary task and a secondary task. In the primary task, participants should follow the guiding line on the ground to drive. The secondary task involves recognizing the shape of the icon in the caution sign and pressing the corresponding button on the steering wheel.

\paragraph{Attention-Shifting Methods}
In our study, we employed three Attention-Shifting Methods as conditions: Sound, Visual and Both. The Sound condition employed spatial sound at the target position to indicate its location. The visual condition used a directional lines to direct participants' attention during the task. Both conditions integrated both visual and Sound cues for participants during the task. We describe the design details in the next section.

\begin{figure*}
    \centering
    \includegraphics[width=\textwidth]{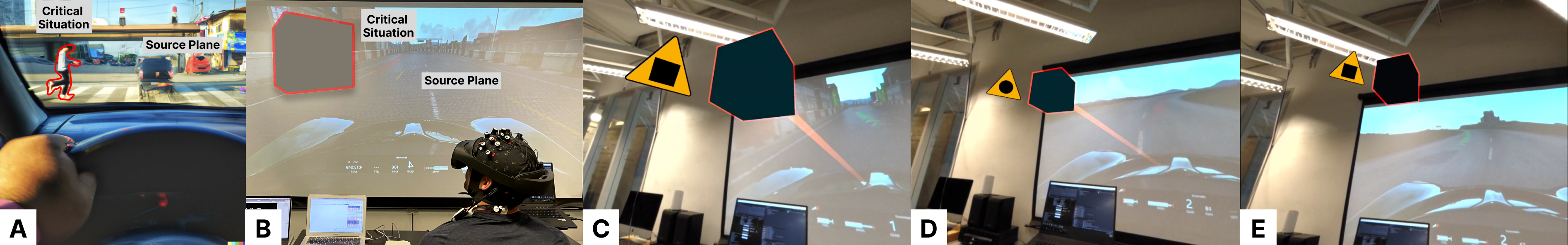}
    \caption{A) driver's actual field of view in the real-world scenario B) Simulating an abstract scenario to capture the driver's response to critical situations in our study C) Visual and Both condition in participant view. D) Circle caution sign icon. E) Sound condition}
    \label{fig:abstractandguidancedesign}
\end{figure*}

\paragraph{Attention-Shifting Methods Design}
Our study highlights the effectiveness of attention shifting methods in mixed reality (MR) by utilizing two novel approaches: spatial sound and directional lines. Spatial sound allows for a more immersive experience by providing auditory cues that are spatially aligned with virtual objects or events in the MR environment. This enhances the driver's ability to locate and attend to specific sources of information. Additionally, the use of directional lines provides visual cues that guide the driver's attention towards critical objects or areas of interest. These advantages of MR, namely the integration of spatial sound and directional lines, have not been previously observed in traditional methods. 
In our study, we simulate an abstract scenario to capture the driver's response to critical situations. The driving view was divided into the source plane and target. The source plane represented the driver's focal point, while the target depicted critical objects like obstacles and pedestrians[Fig.\ref{fig:abstractandguidancedesign}A-B]. We used red outlined squares and a caution sign to inform the driver of the immediate response required to deal with the critical situation. The caution sign contains two icons, a square and a circle, representing different types of critical situation event messages. To draw attention without disruption, we generated a transparent orange line connecting the source plane's center to the target's center in the driving scenario. [Fig.\ref{fig:abstractandguidancedesign}C-E]

\paragraph{Apparatus}

\begin{figure*}[h]
    \subfigure{\includegraphics[width=0.35\textwidth]{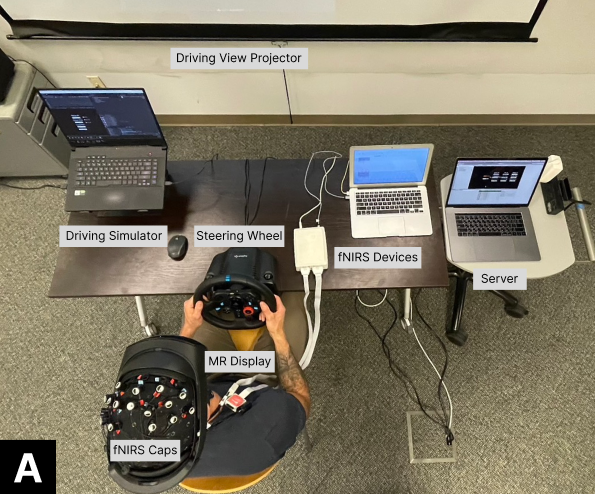}} 
    \subfigure{\includegraphics[width=0.35\textwidth]{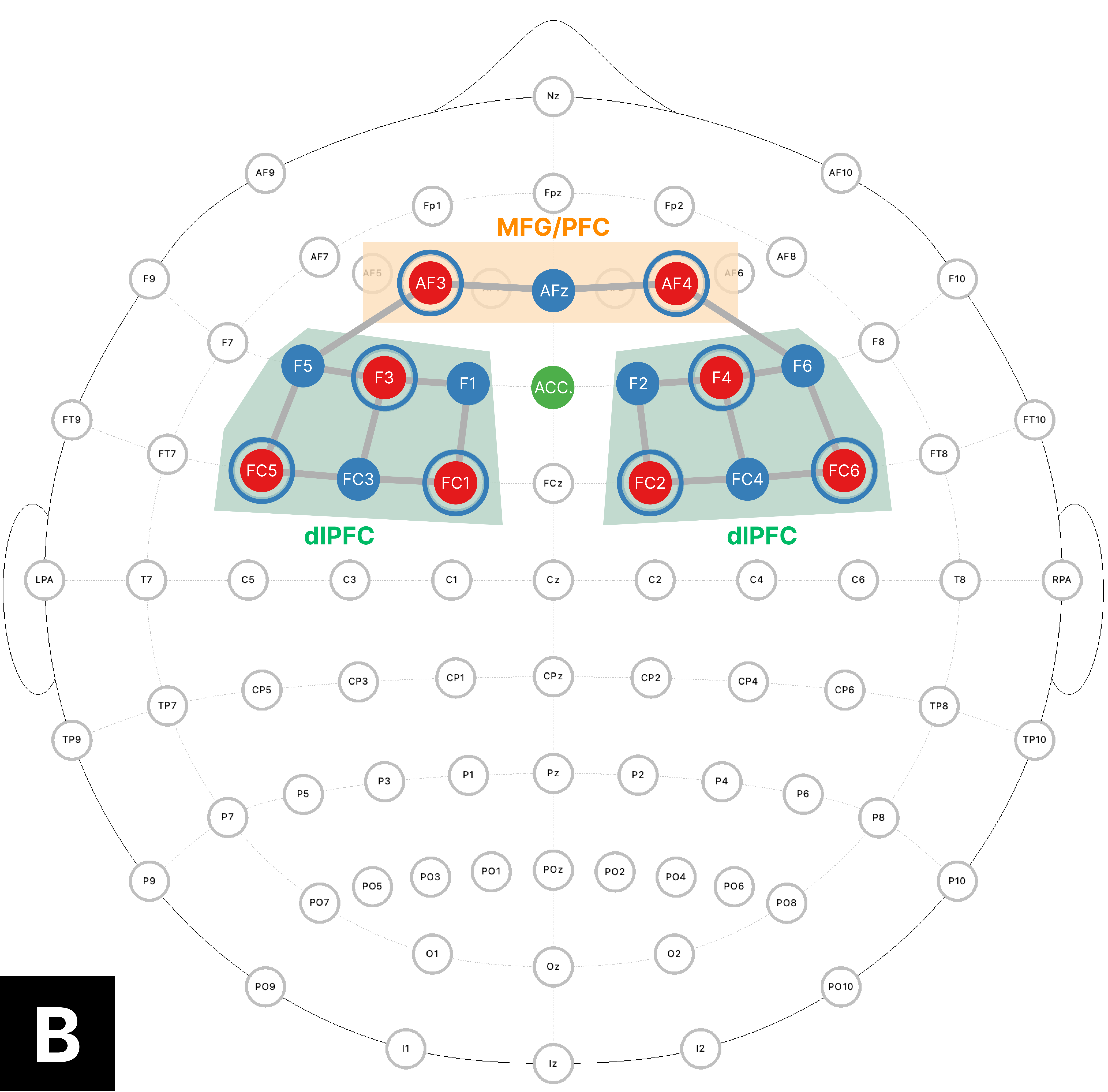}} 
    \caption{A) Study Apparatus B) This images shows the montage used for the experiments in this study. Highlighted are the regions of interest. Blue shows the channels measuring the bilateral dlPFC while orange indicates the channels measuring activity in the MFG/PFC}
    \label{fig:appartausandmontage}

\end{figure*}

For the driving simulator setup, we used a laptop with an Intel i7-10700k CPU, RTX 2080 GPU, and 32GB RAM. Participants controlled the vehicle with a Logitech G29 steering wheel. The game view was projected on a 90-inch screen using an Epson projector. For fNIRS measuring, we utilized the NIRx NIRSport2 with 8 sources and 8 detectors. The Quest Pro MR display provided targets, notifications, and eye tracking data. A Macbook Pro served as the server, using OSC to synchronize signals and receive eye tracking data. [Fig.\ref{fig:appartausandmontage} A] The chosen track was the highlands drift in Assetto Corsa, with the Lotus Elise SC as the vehicle for improved control.

\subsection{Measurement}
\paragraph{fNIRS Montage Design for MR Device}
Given the use of an MR headset, we customized the standard prefrontal cortex measurement montage to ensure compatibility and adequate coverage of our region of interest (ROI), particularly the dorsolateral prefrontal cortex (dlPFC) and medial frontal cortex (MFG). Our new cap design, inspired by an effective montage\cite{zapala_recognition_2022}, balanced capturing cortical activity with comfort under the Quest Pro device. These brain regions are known for their involvement in task switching and higher-order thinking\cite{hyafil_two_2009, japee_role_2015}. Our primary region of interest for these tasks was the prefrontal cortex for its role in executive function and decision making.[Fig.\ref{fig:appartausandmontage} B] Within the frontal cortex our sub-regions of interest were the bilateral dorsolateral prefrontal cortex (dlPFC) for it's roles in task switching as well as the middle frontal gyrus (MFG) for it's role in processing external stimuli during attention shifting.\cite{brass_role_2005,japee_role_2015}. EEG 10/10 landmarks on a NIRx fNIRs cap were used to guide the channel placement for our regions of interest.

\paragraph{NASA-TLX}
To assess participants' subjective mental workload, we utilized the NASA-TLX survey. This survey was administered immediately after each condition and provides a numerical rating on a scale of 0 to 100, indicating the level of mental workload experienced. It considers six sub-scales: mental demand, physical demand, temporal demand, effort, performance, and frustration level. These sub-scales are weighted and combined to produce a single value representing the overall cognitive load. This value was utilized in our analysis of participants' subjective mental workload during the study \cite{prabaswari2019mental}.

\paragraph{Behavioral Metrics}
In our study, we measured two metrics: response time and error rate. Response time was measured from the target appearance to participant's response. Error rate was calculated based on incorrect reactions. For instance, participants should press the circle button for caution sign icons is circle shape. Pressing the square button would be considered a wrong answer.

\subsection{Procedure}
The experiment consisted of four phases, with total lasting approximately an hour for each participant. The phases included: (1) introduction, (2) signal optimization, (3) practice drive, and (4) experimental drive. During the introduction, participants were briefed about the study and then proceeded to wear the fNIRS caps for signal optimization. Once the signal was confirmed, participants put on the MR display and began practicing driving while familiarizing themselves with the button responses. In the experimental drive phase, participants were instructed to drive on the track, with a target appearing every 5 seconds after each response. Each condition comprised 4 blocks, with each block lasting 60 seconds and a 20-second interval in between. After completing a condition, participants were asked to answer the NASA-TLX questionnaire and had a 2-minute rest period. The order of conditions was fully counterbalanced throughout the experiment. Finally, participants concluded the study by answering questions about their experience.

\subsection{Data Processing}
All fNIRs data processing was done using the open-source Python library MNE version 1.4.2. Our study was based around a within group block design delineated by event triggers in the fNIRs data stream. The triggers used enabled ease of separation between the "Sound", "Visual" and "Both" conditions. The events were converted into epochs of 51 seconds from the time of the trigger. 

\subsubsection{Pre-processing}
During the data pre-processing we used scalp coupling index (SCI) in order to validate the quality of the optode. We decided to reject any channel that resulted in an SCI below 0.8. The data was then re-sampled to 0.8 Hz. We then used temporal derivative distribution repair (TDDR) in order to correct for baseline shifts and spike artifacts. The raw hemodynamic data was then passed through both a low-pass filter of 0.02 and a high-pass filter of 0.3. The low-pass and high-pass filters were selected for their ability to remove the physiological noise such as heart beat from the sample data. The bandpass filters used during data pre-processing were values recommended by the MNE documentation as default values.

\subsubsection{Data Analysis}
Data analysis was carried out using a two level approach separated into individual and group processing. During individual processing each subjects data is pre-processed as mentioned above. This step also involves the conversion of the data streams into epochs. The epochs were created using event triggers sent from Unity into Aurora to delineate various events. Since this experiment was done as a block design we generated epochs from -1 seconds to 50 seconds from the time of the trigger. These epochs were then aggregated and passed onto the group level analysis portion of the analysis pipeline. The statistical analysis for the group was carried out using a generalized linear model (GLM). In this step we incorporate our short channel measurements as well as accelerometer data into the design matrix as regressors to further remove any physiological noise and motion artifacts from the data stream. The GLM data is then passed through a false detection rate (FDR) test in order to minimize the presence of false positives in our channel significance results.

\subsection{Results}
\subsubsection{Cognitive Work Load}
\begin{figure*}
    \centering
    \includegraphics[width =\textwidth]{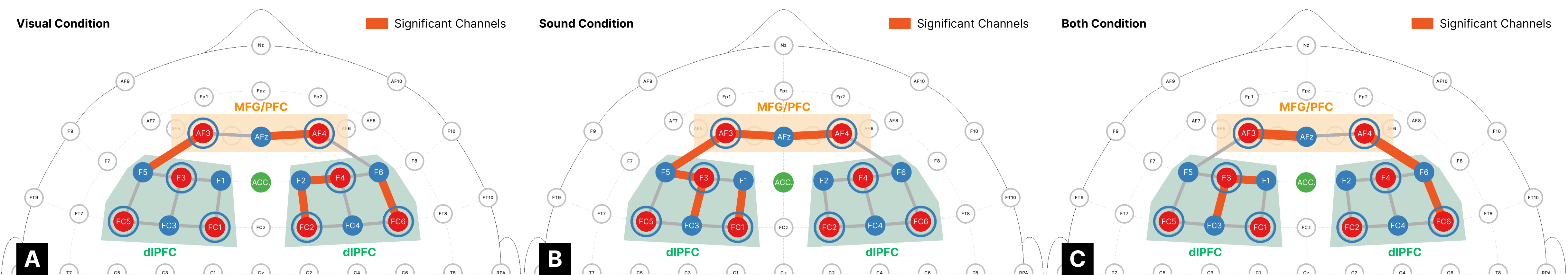}
    \caption{This image depicts the significant channels through a topographical map of the cortical surface. All channels and cortical activation portrayed here survived FDR correction with an alpha set at 0.05.}
    \label{fig:montageresult}
\end{figure*}

\paragraph{fNIRs Results}
In order to process the fNIRs results we relied on both a two-way ANOVA using the python package statsmodels as well as a channel by channel analysis with FDR correction generated by the GLM. [Fig.\ref{fig:montageresult}]
The two-way ANOVA indicated there was a significant effect between the channel and beta values produced during GLM from the hemodynamic reading 
\textit{(p < 0.001, df = 35, SSE = 2.12e-10, F = 2.85)} as well as a significant interaction effect between the channel and condition on those beta values \textit{(p < 0.001, df = 70, SSE = 5.84e-10, F = 3.92)}. The channel by channel analysis further demonstrated the significance of the channel to condition interactions.

\paragraph{Interpretation of Significant fNIRS Channels}
The significant channels seen in every condition can be found in Fig. \ref{fig:montageresult}. While the results are still preliminary, they do show that participants had higher PFC activation with "Sound" condition than the other two conditions. The right dlPFC also seems to be preferentially activated when participants were in the "Visual" condition. When participants were in the "Both" condition it seems that the effects were more even spread around the frontal cortex. It is interesting to note that the effects on cognitive workload in the "Both" condition did not seem to be a result of an additive effect of the visual and sound condition.

\paragraph{NASA-TLX Weighted Ratings} In this analysis we concerned ourselves mostly with the NASA-TLX weighted rating for it's ability to distill cognitive work load into a single number. The "Sound" condition responses produced a weighted rating ranging from 23.33 to 74.00 (M = 56.52, SD = 18.91). The "Visual" condition responses produced a weighted rating ranging from 45 to 66 (M = 56.43, SD = 7.91). Finally, the "Both" condition produced a weighted rating ranged from 35.33 to 77.33 (M = 56.05, SD = 15.78).

\paragraph{Statistical Evaluation of NASA-TLX}
We used a one-way ANOVA to evaluate the statistical relevance of the NASA-TLX results. From a high level, the Weighted Rating showed no statistical significance between the conditions \textit{(p = 0.998, df = 2, F = 0.002)}. Among the remaining six sub-scales neither sub-scale showed a statistically significant difference between the three conditions.

\subsubsection{Behavioral Metrics}
\paragraph{Target Finding Error}
We used a two-way ANOVA to examine the impact of condition and target area on target identification error rates. The results showed that both condition and target area did not have a significant effect on error rates. The error rates for the "Visual" condition were 14.57\%, the "Sound" condition error rate was 10.14\%, and the "Both" condition error rate was 14.97\%.

\paragraph{Target Finding Time}
The target finding time varied across the conditions, with the "Visual" condition having a mean of 3.23s \textit{(SD = 2.82s, N = 151)}, the "Sound" condition having a mean of 3.60s \textit{(SD = 5.74s, N = 138)}, and the "Both" condition having a mean of 2.87s \textit{(SD = 3.05s, N = 147)}. The two-way ANOVA results indicated that neither the "target area" (F = 1.0380, p = 0.3089) nor the "condition" (F = 0.7818, p = 0.4582) had a statistically significant effect on the target finding time.

\section{Discussion}
\subsection{The Relation Between fNIRS and NASA-TLX}
While both fNIRS and NASA-TLX have advantages and limitations, relying solely on NASA-TLX may not offer a comprehensive understanding of cognitive workload across tasks. NASA-TLX is cost-effective but provides only a snapshot of subjective experiences without uncovering underlying reasons. Conversely, fNIRS enables real-time measurement of physiological responses, shedding light on underlying mechanisms behind behavioral patterns. Our study found significant differences in cortical responses and task-based behaviors not fully captured by NASA-TLX alone. Integrating these modalities allows researchers to approach HCI questions from multiple perspectives, uncovering valuable insights overlooked by a single metric or test. This integration offers a more nuanced understanding of cognitive processes in HCI tasks.

\subsection{Target Presentation Intervals}
One limitation of our study is the fixed 5-second time interval between target displays during the experimental drive phase. We chose this interval to control the hemodynamic response, aligning with fNIRs block design practices. However, it may increase hazard predictability, reducing ecological validity. In future research, we aim to address this limitation by minimizing event predictability, offering participants a more realistic driving experience.
\section{Limitation and Future Work}
\subsection{Exploring Limitations and Future Possibilities of MR Experiments in Driving Scenarios
}

In our study, we explored limitations when using MR devices in experiments. Current MR devices have constraints in FOV, resolution, and comfort, impacting user experience. Light variations in different environments can affect object perception and integration, reducing realism. These limitations, like brightness issues, may be resolved as MR technology advances. We successfully integrated MR display and fNIRS measuring in a driving simulator. Future plans involve evaluating our study in real cars, enabling deeper insights and comparisons with simulator results to advance understanding.

\subsection{Using neural network model to predict driving events }
In future research, we aim to explore leveraging machine learning and deep learning techniques to automatically label driving events based on users' hemodynamic responses. This eliminates the need for manual triggering and allows for a more comprehensive dataset. As a preliminary step, we propose training a support vector machine (SVM) model using event triggers obtained from this experiment. If successful, our next step would involve training a transformer model using fNIRS data and video footage of participants' driving. Achieving high accuracy with this approach would provide a valuable tool for studying automotive user interfaces in the HCI community.
\section{Conclusion}
This study investigated drivers' attention allocation in critical situations using a mixed-reality display. A driving simulator with a tailored fNIRS montage was used to explore the impact of different notification types and positions on cortical responses and participant performance. During analysis we found significant differences in fNIRS channel activation between conditions. Despite the fNIRs results we found no further significant differences in subjective reports and behavioral metrics between conditions. Certain metrics such as the interaction between condition and target area on the target finding precision were close to reaching significance \textit{(P = 0.051, SSE = 0.69, df = 2.0, F = 2.99)}. This suggests that further data collection would be beneficial to further our understanding of the effect between these interactions. 
Future plans involve gathering real-world data and employing machine learning and deep learning techniques to predict events based on users’ brain activity. 

\bibliographystyle{ACM-Reference-Format}
\bibliography{reference}

\end{document}